%% file: CMOS_Paper.tex
\pdfoutput=1 
\documentclass{JINST}
\usepackage{graphicx}
\usepackage{subfigure}
\usepackage{textcomp}
\usepackage{upgreek}
\usepackage{rotating}
\usepackage{amsmath}
\usepackage{xspace}

\newcommand{\ToT}{\ensuremath{\textsf{To\kern -0.1em T}}\xspace}
\newcommand{\lvl}{\textsf{Lvl1}\xspace}
\newcommand{\FDAC}{\textsf{FDAC}\xspace}
\newcommand{\thrv}{\textsf{Th}\xspace}
\newcommand{\VNFB}{\textsf{VNFB}\xspace}
\newcommand{\VNOut}[1]{\textsf{VNOut#1}\xspace}

\title{Subpixel Mapping and Test Beam Studies with a HV2FEI4v2 CMOS-Sensor-Hybrid Module for the ATLAS Inner Detector Upgrade}

\author{T. Bisanz, J. Gro\ss e-Knetter, A. Quadt, J. Rieger,  J. Weingarten \\
II. Physikalisches Institut, Georg-August-Universit\"{a}t G\"{o}ttingen\\
Friedrich-Hund-Platz 1, 37077 G\"{o}ttingen, Germany\\
E-mail: \email{jens.weingarten@uni-goettingen.de}}

\abstract{The upgrade to the High Luminosity Large Hadron Collider will pose unprecedented challenges to the tracking systems
of all experiments. Recent advancement of active pixel detectors designed in CMOS processes provide attractive alternatives to the 
well-established hybrid design using passive sensors. 

This article presents studies with a high-voltage CMOS active pixel sensor designed for the ATLAS tracker upgrade.
The sensor is glued to the read-out chip of the Insertable B-Layer. The device under test is structured in smaller pixels than 
those of the attached read-out chip with information on the hit CMOS pixel encoded 
in the output signal (subpixel encoding). Test beam measurements mapping recorded hits in the read-out chip to CMOS pixels,
thus validating the subpixel encoding, are described.}


\begin{document}


\input{sections/genintro}

\input{sections/cmosintro}

\input{sections/parstud}

\input{sections/tbsetup}

\input{sections/subpix}

\input{sections/resolution}

\input{sections/hiteff}

\input{sections/conclusion}

\acknowledgments

One of the authors wants to thank the Konrad-Adenauer-Stiftung PhD scholarship program. The work was supported by the Federal Ministry of Education and Research (BMBF), FSP103 under contract number 05H15MGCAA and FIS under contract number 05H15MGCA9. The measurements leading to these results have been performed at the Test Beam Facility at DESY Hamburg (Germany), a member of the Helmholtz Association (HGF).

\bibliographystyle{Bib/atlasnote}
\bibliography{Bib/Bibliographie}

\end{document}

%% file: sections/genintro.tex
\section{Introduction}

The Large Hadron Collider (LHC) is scheduled to be upgraded to the High Luminosity Large Hadron 
Collider (HL-LHC) from 2023 onwards. After this Phase-II Upgrade, the instantaneous peak luminosity is 
expected to reach values of up to  $7.5 \times 10^{34} \,\mathrm{cm}^2\mathrm{s}^{-1}$.
The challenges resulting from expected unprecedented hit rates and radiation levels in the innermost part
of the ATLAS detector will 
be met by a full replacement of the current ATLAS Inner Detector (ID)
with an all-silicon tracking system, the ATLAS Inner Tracker (ITk)~\cite{HL-LoI,HL-SD}. 

The ITk foresees a pixel detector as its innermost component that will cover an area of approximately 10~m$^2$, corresponding to about
ten times that of the pixel detector of the ID. More cost-effective technologies
than the hybrid pixel module concept connecting sensor and read-out chip pixel
by pixel with bump bonds are therefore attractive.

Industrial CMOS processes offer design features resulting in a high breakdown voltage which allows
the creation of a deep depletion zone suitable for detection of charged particles. CMOS processes
are typically applied to wafers larger than those used for passive, high-resistivity bulk sensors.
In combination with cost-effective interconnection methods, such as wafer to wafer bonding or gluing techniques,
active CMOS pixel sensors are a promising candidate for future applications. In addition, CMOS designs allow to
realise smaller pixel sizes than would be possible with devices designed for bump bond interconnection,
which is beneficial in a high track density environment.

This report presents parameter studies and test beam measurements specifically addressing a new concept 
of reading out several smaller CMOS pixels connected to one pixel on the read-out chip
to benefit from improved resolution and track separation.
This paper is based on the results and contains sections from Ref.~\cite{Rieger_Thesis}.

%% file: sections/cmosintro.tex
\section{Active High-Voltage CMOS Sensors}

Processes such as HV-CMOS offer new approaches to realise active pixel detectors capable of coping 
with rate and radiation levels expected
at the HL-LHC. The high-voltage technology allows a combination of standard low-voltage CMOS transistors, 
which are used to implement internal electronics, and the bias of the sensor with more than -50~V. 
Consequently, a depleted zone of $10\, \upmu\mathrm{m}$ to $100\,\upmu\mathrm{m}$ is formed, depending on the CMOS process and
the bulk resistivity. The charge carriers released by ionising particles in the depleted area are
collected by drift to allow a fast read-out that fulfils the HL-LHC requirements.

The HV-CMOS technology is applied in a large variety of CMOS processes, which are industrially available 
from several vendors. This report describes studies with the HV2FEI4v2~\cite{IvanHVCMOS} which is implemented
in the Austria Microsystems (AMS) 180 nm high-voltage CMOS process~\cite{Ivan2007}. It is glued to
the FE-I4 readout chip~\cite{FEI4B_2012, FEI4B_2013}, forming a capacitively coupled pixel detector (CCPD)~\cite{IvanCCPD}.

The FE-I4 contains read-out circuitry for 26~880 pixels arranged in 80 columns
of $250\,\upmu\mathrm{m}$ pitch by 336 rows of $50\,\upmu\mathrm{m}$ pitch. 
Each FE-I4 pixel is composed of an amplification stage followed by a discriminator with an independently adjustable
threshold. For each hit passing the discriminator threshold, the time at which the threshold is crossed is recorded
along with the 4~bit Time over Threshold (\ToT), measured in units of 25~ns. The \ToT is defined as the time that the 
amplifier output signal remains above threshold. The correspondence between collected charge and \ToT can 
be adjusted individually for each cell by the feed-back current of the amplification, controlled by a pixel register DAC (\FDAC).
Information from all discriminators which detected a hit is kept in the chip for a programmable latency interval  
after which the information is sent on the output link if a trigger is supplied. 
Hits from up to 16 consecutive LHC clock cycles (25~ns) are sent upon trigger reception with their
respective timing noted by a trigger counter (\lvl).

\begin{figure}[tbph]
\centering
\includegraphics[width=0.6\textwidth]{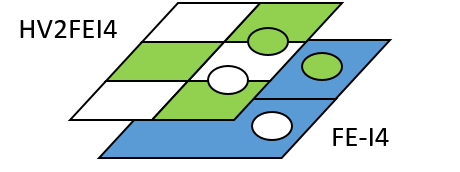}
\caption{Schematic drawing of the HV2FEI4v2 subpixel structure of one unit cell. The three white (light) sensor 
pixels are connected to the FE-I4 readout chip pixel with the white pad and the green (dark) ones to the FE-I4 pixel 
with the green pad.}  \label{fig:subpixel}
\end{figure}

The HV2FEI4v2 sensor prototype has a size of approx.~$2.2\times 4.4\,\mathrm{mm}^2$, which is much
smaller than the FE-I4. The sensor pixels are arranged in 240 unit cells, each containing six pixels. 
Three sensor pixels are connected to one FE-I4 readout chip pixel. The schematic drawing in 
Figure~\ref{fig:subpixel} shows the subpixel mapping for one unit cell.
The standard pixel cell consists of an amplifier, the corresponding feedback circuit and a comparator.
The working points of the transistors are set by external bias voltages, most importantly the external threshold 
voltage, \thrv. Other parameters are controlled by global and pixel registers.
The global register contains settings for 19 DACs and six switches. The adjustment of the
feedback current is controlled by the \VNFB DAC. The parameters \VNOut{1}, \VNOut{2}, and
\VNOut{3} adjust the amplitude of the output signal of the three different subpixels.

%% file: sections/parstud.tex
\section{Parameter Studies}

\subsection{Read-out System and Injection-based Measurements}\label{sec:usbpix}

Data acquisition and control of both the HV2FEI4v2 and the FE-I4 of the CCPD hybrid under study were
carried out with the USBpix system~\cite{Backhaus_USBpix}. Operation of FE-I4 chips (without CMOS sensors) 
with USBpix was fully implemented in the course of module tests for the Insertable B-layer (IBL)~\cite{IBL}.

While communication between USBpix and FE-I4 relies on the programmable digital electronics
of the USBpix main board (Multi-IO), a General Purpose Adapter Card (GPAC) connected to the Multi-IO board
provides additional DC and slow input signals needed to control the HV2FEI4v2. Both
boards are steered by a C++ software package that permits simultaneous control and read-out
of HV2FEI4v2 and FE-I4 as would be required during ATLAS data taking and calibration~\cite{Rieger_USBPix}.

Test and calibration of the HV2FEI4v2 are performed using an internal injection mechanism:
a well-defined charge is injected directly into the
analogue part of all pixels of the HV2FEI4v2 at the same time via the GPAC injection
circuit. The generated signal is then read out by the FE-I4. With a measurement injecting
a charge well above the expected HV2FEI4v2 discriminator threshold, the capacitive coupling of HV2FEI4v2 
to FE-I4 is validated (\textit{Analog Test}). A similar measurement records and possibly adjusts the \ToT measured 
by the FE-I4. A \textit{Threshold Scan} is based on injections, automatically varying the 
injection voltage from the GPAC during the scan. The injection is repeated several times and
the number of hits per injection voltage as measured by the FE-I4 is recorded. During the scan, the injection
voltage is varied from 0~V to 1.5~V in 61 steps. The resulting number of hits as function of the injection voltage
is fitted with an error function. The fit determines the injection voltage at the comparator 
threshold (threshold equivalent voltage, Th$_{\textrm{EV}}$) as the voltage at 50\% of the occupancy 
while the fit parameter describing the spread of the turn-on behaviour
corresponds to the equivalent noise of the HV2FEI4v2 analogue cell.

\subsection{HV2FEI4v2 Comparator Threshold}\label{sec:compthresh}

\begin{figure}[tb]
\centering
\subfigure[External threshold voltage \thrv and \VNFB DAC.\label{fig:Vthr+VNFB}]{\includegraphics[width=0.48\textwidth]{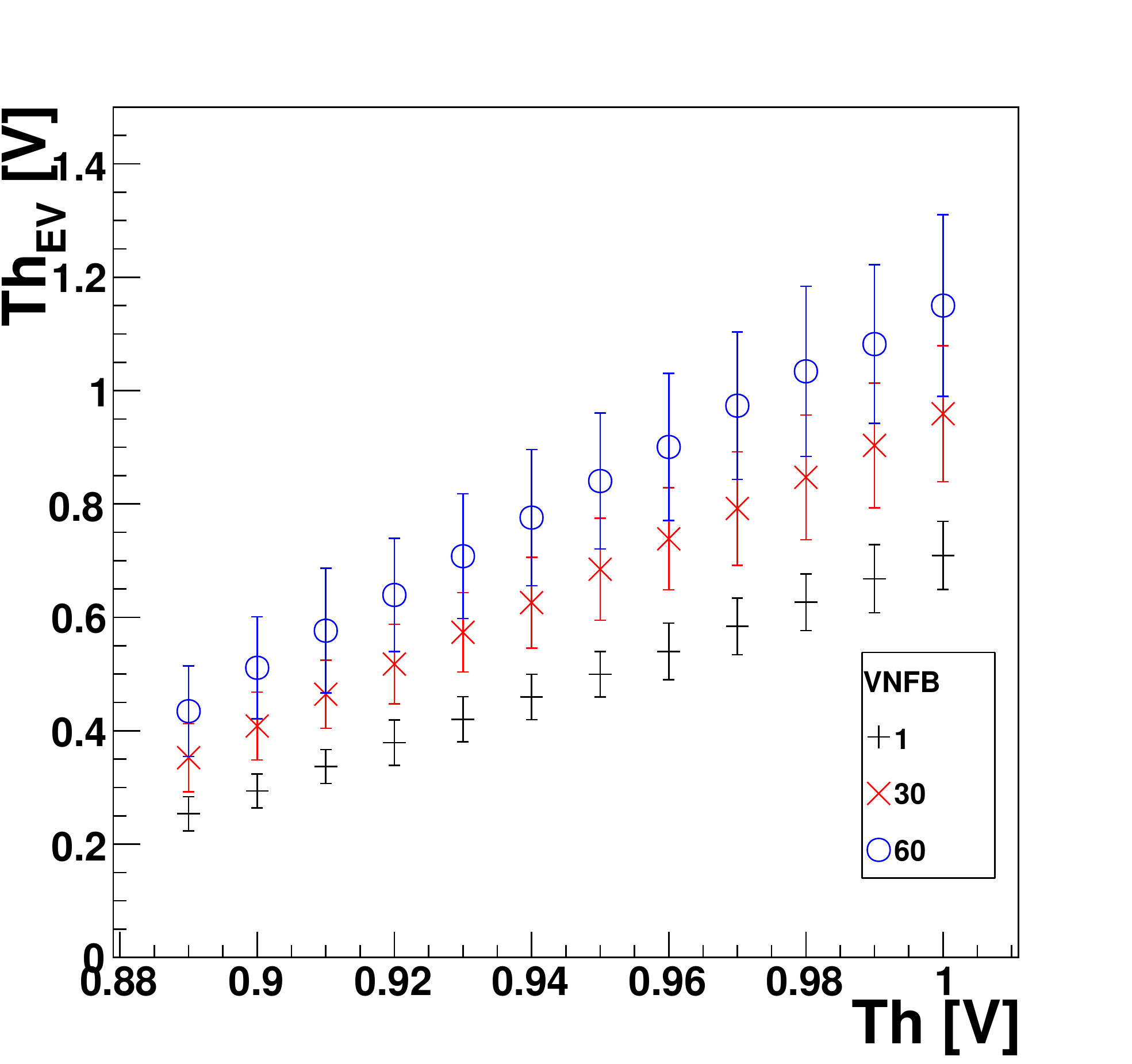}}
\subfigure[\VNOut{} DAC.\label{fig:Thr_VNOut}]{\includegraphics[width=0.48\textwidth]{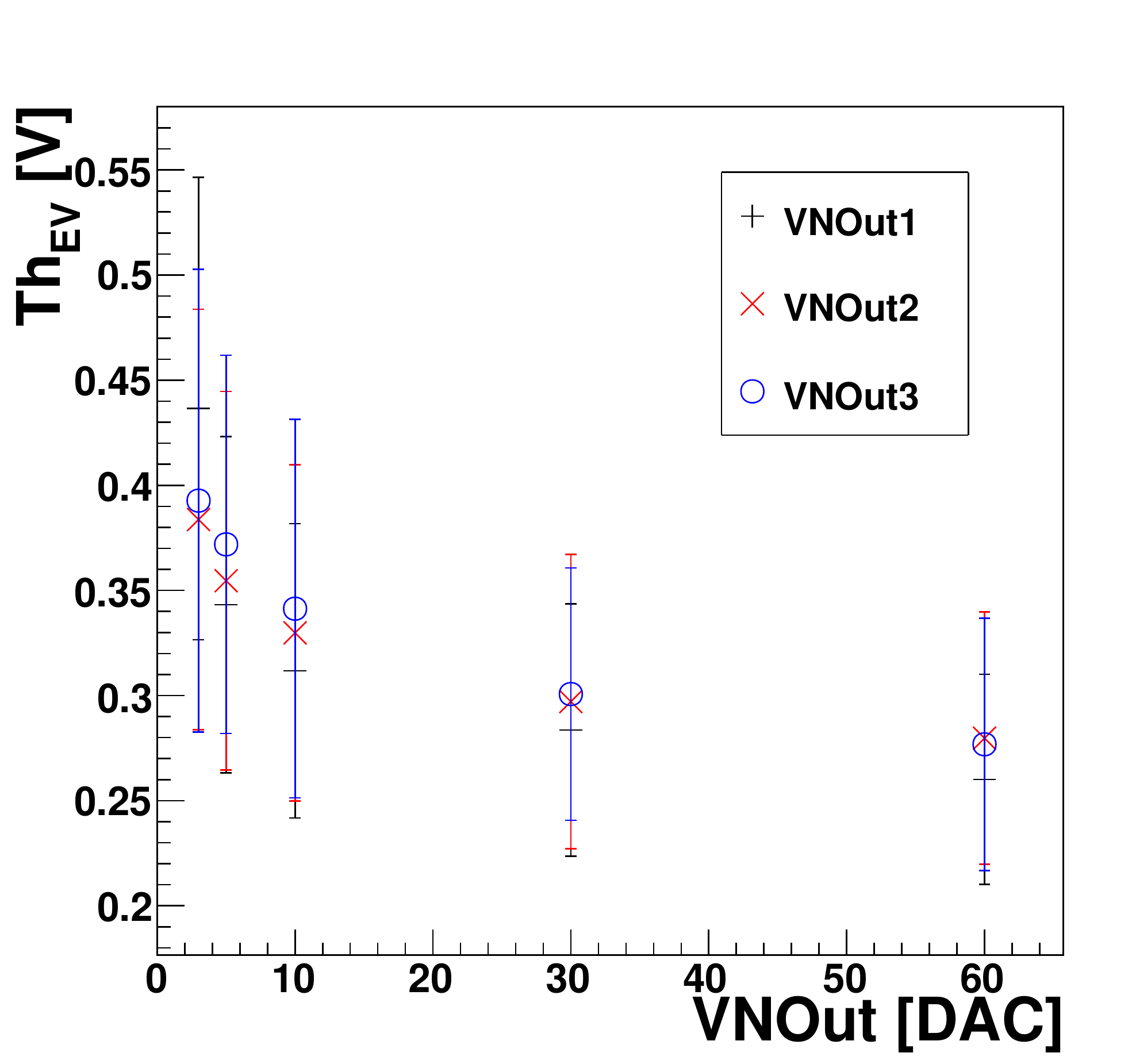}}
\caption{Threshold equivalent voltage, Th$_{\textrm{EV}}$, of the HV2FEI4v2 as measured with a \textit{Threshold Scan} for different parameter settings. The error bars show the pixel dispersion.}
\label{fig:Vthr}
\end{figure}

The external threshold voltage \thrv is assumed to be the parameter that influences the comparator threshold Th$_{\textrm{EV}}$
the most. Figure~\ref{fig:Vthr+VNFB} shows the results of different \textit{Threshold Scans} for external threshold 
voltages between the minimum stable value\footnote{For voltages below 0.89~V no measurement
was possible, because for some pixels the threshold was so low that only noise hits were
registered.} 0.89 V up to 1 V.
The error bars indicate the pixel dispersion. Both Th$_{\textrm{EV}}$ averaged over all pixels and the corresponding dispersion
increase with increasing external threshold voltage. 

In addition, the influence of the feedback parameter \VNFB was studied by measuring Th$_{\textrm{EV}}$ for
three \VNFB DAC settings.
\VNFB has a stronger influence on Th$_{\textrm{EV}}$ than the external threshold voltage. For large feedback 
currents the amplitude of the amplifier output signal is reduced, because part of the signal charge is compensated. 
This effect is known as ballistic deficit and leads to an increase of the effective comparator threshold at high values 
of \VNFB.
Furthermore, the dispersion of the distribution increases with \VNFB. This effect results from the fact that the 
feedback transistors suffer more from process fluctuations than e.g. the comparator transistors.

Because the \VNOut{1/2/3} DACs will play an important role in the subpixel mapping, their influence on the 
comparator threshold was investigated. Figure~\ref{fig:Thr_VNOut} shows the measurement of the injection voltage 
at the threshold as function of the \VNOut{} DAC values for all three subpixels. The error bars again indicate 
the pixel dispersion. The measurement was performed at the minimum stable external threshold voltage of 0.89~V. In the 
measurement for \VNOut{1} only the subpixel type~1 was enabled, as for the measurements with \VNOut{2} 
and \VNOut{3} only the subpixel types~2 and~3, respectively.
The \VNOut{1/2/3} DACs influence the threshold equivalent voltage. This effect was not expected, because the 
\VNOut{1/2/3} parameters are supposed to only influence the signal amplitude after the comparator. No significant difference 
among the three subpixels is measured.

\subsection{Tuning Towards Subpixel Mapping}\label{sec:subpixtune}
For a successful subpixel decoding, the \ToT response of the FE-I4 needs to be adjusted to the amplitude of the sensor 
signal~\cite{MalteSubPix}. Before tuning the signals of the subpixel types 1, 2 and 3, the impact of the \VNOut{1/2/3} 
DACs on the \ToT is investigated.

\begin{figure}[tbph]
\centering
\hfill
\begin{minipage}[t]{.45\textwidth}
  \centering
  \includegraphics[width=\textwidth]{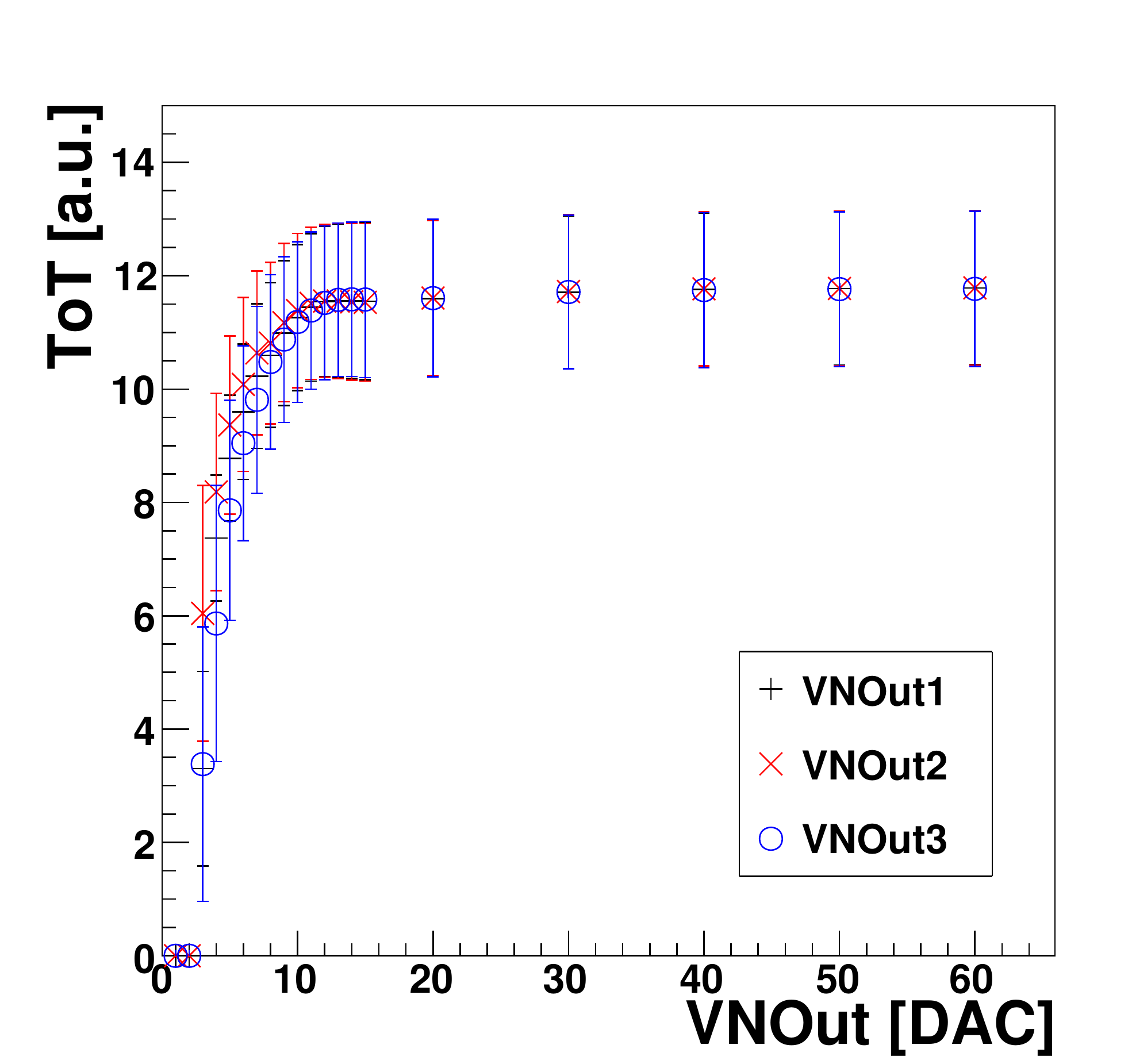}
  \caption{\ToT response as a function of \VNOut{1/2/3} DACs for the three subpixel types for a given feedback tuning of the FE pixels. 
           The error bars show the standard deviation of the distribution.}
  \label{fig:VNOut+ToT}
\end{minipage}%
\hfill
\begin{minipage}[t]{.46\textwidth}
  \centering
  \includegraphics[width=\textwidth]{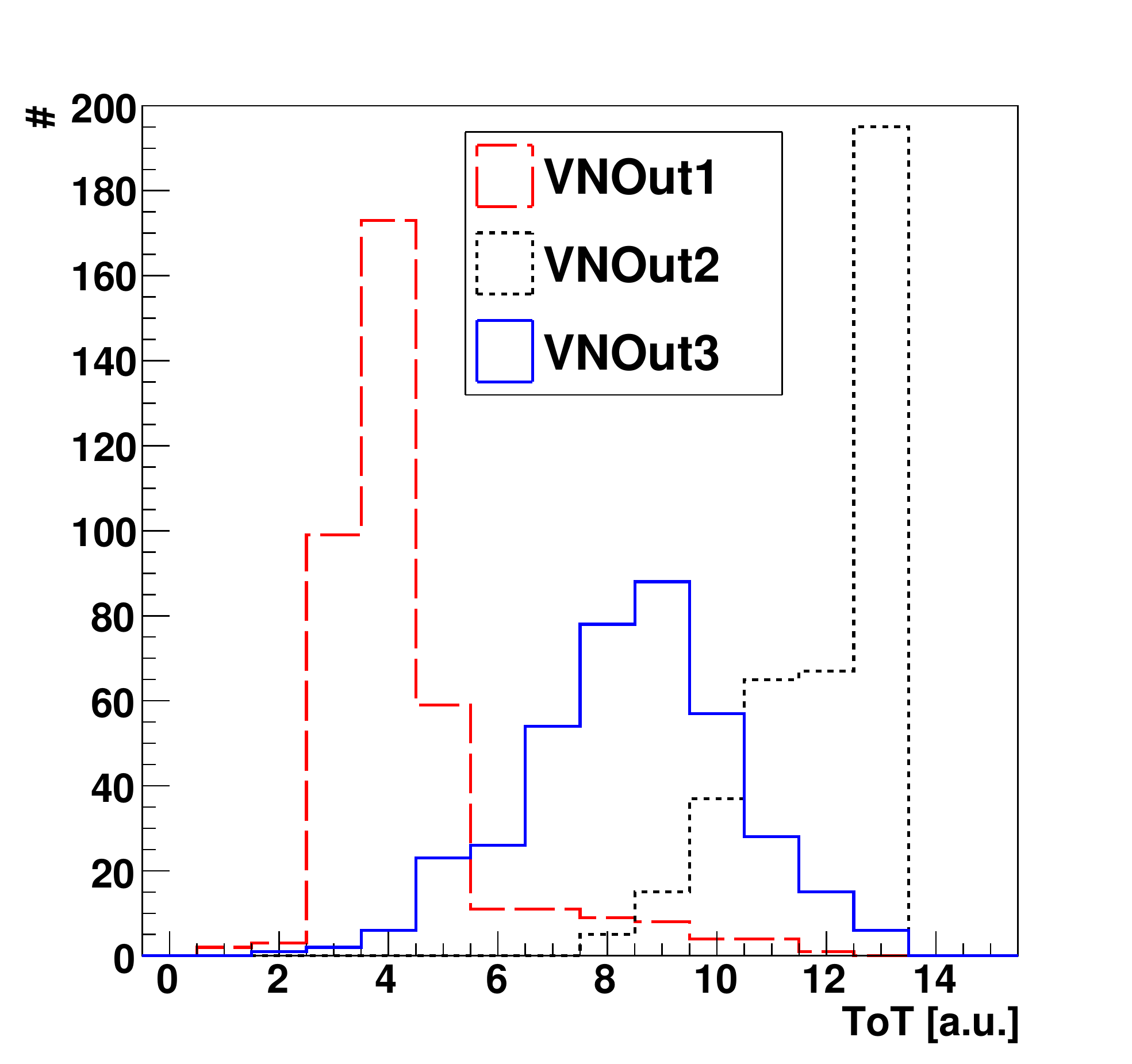}
  \caption{\ToT distribution of the HV2FEI4v2 module for the three subpixel types measured in three separate \textit{Analog Test} scans.\protect\\}
  \label{fig:HVToT}
\end{minipage}
\hfill
\end{figure}

Only one subpixel type was enabled at a time and the \ToT response to different \VNOut{} settings was measured. 
The same was repeated for all three subpixel types. Figure~\ref{fig:VNOut+ToT} shows the \ToT response as a function of 
\VNOut{1/2/3} DAC values.
\ToT saturates for \VNOut{} DAC settings above 12. 
In this limited dynamic range the output signal of the different subpixel types varies by up to 2 \ToT units with \VNOut{2} having the highest \ToT response. 

Considering the result above, the \FDAC of the FE-I4 readout chip was adjusted to the sensor response. The 
\FDAC values were tuned to produce a \ToT value of 3 for a signal with \VNOut{1} set to 3 to get the lowest possible \ToT response. 
To get the largest possible separation between the \ToT response 
of the three subpixels, \VNOut{2} was set to 60 and \VNOut{3} was set to 5.

With these \VNOut{1/2/3} settings an \textit{Analog Test} was performed to measure the average \ToT response per pixel. 
Figure~\ref{fig:HVToT} shows the \ToT distributions for the three subpixels after the tuning. Most of the 
subpixels can be separated, but there is some overlap between the distributions.

%% file: sections/tbsetup.tex
\section{Test Beam Measurements}	\label{sec:tbsetup}

\subsection{Set-up}

A EUDET-type beam telescope~\cite{EUDET} was used to reconstruct the tracks of the particles passing through the assembly under test. Thus, hit information from the assembly (hit position, timing) can be compared to the extrapolated track position at the position of the assembly, allowing the calculation of spatially resolved hit efficiency and the resolution of the assembly.

The high resolution EUDET-type beam telescope is based on monolithic active pixel sensors. It consists of six planes of Mimosa26 sensors, the positions of which can be adjusted to optimise the tracking resolution. The pixel pitch is 18.4~$\upmu$m, and the single-point resolution of each plane is below 4~$\upmu$m. Data acquisition is provided by the EUDAQ framework~\cite{EUDET2016}. Scintillators placed in front of the telescope provide track triggers, while a planar silicon pixel detector, read out by an FE-I4 chip allows the definition of a trigger region-of-interest, adapted to the physical dimensions of the HV2FEI4v2. Operation of HV2FEI4v2 and FE-I4 was
realised with the USBpix system described in Section~\ref{sec:usbpix}. The existing USBpix software provides full integration with the EUDAQ framework.
The integration time of the beam telescope is 115.2~$\upmu$s, while the assembly under test integrates for 400~ns. 
Therefore, a second FE-I4 module is installed as a timing reference. 

Data was taken at test beam line 21 at the DESY II accelerator. This beam line provides 1-6~GeV electrons. For the measurements presented here, a beam energy of 5~GeV was used to achieve a high beam intensity, leading to a trigger rate of approx.~1~kHz.

To take data at the DESY II test beam, the assembly under test, together with the USBPix readout hardware, was mounted on a custom made mechanical holder between the third and fourth planes of the beam telescope. The sample was oriented perpendicular to the beam direction. The planes of the telescope were arranged to optimise track resolution given the level of multiple scattering expected at the particular beam energy.

\begin{figure}[t]
\centering
\includegraphics[width=0.6\textwidth]{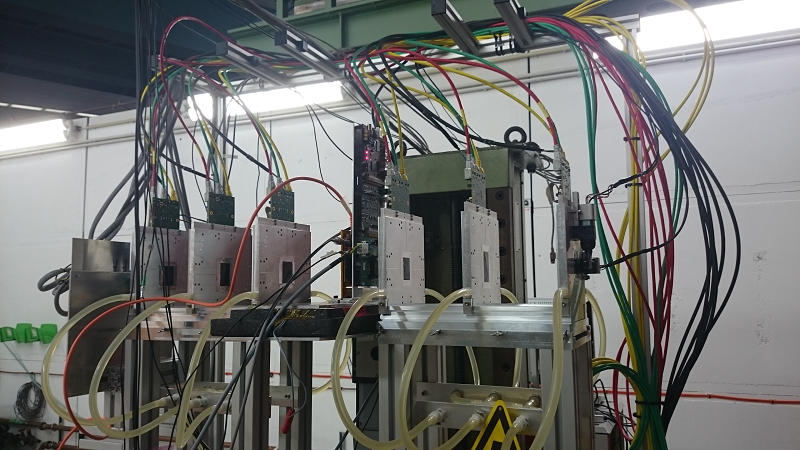}
\caption{The DESY II test beam setup. Shown are the CCPD with the FE-I4 reference plane, together with the six planes of the telescope.}  \label{fig:TB_foto}
\end{figure}

The coordinate system for the test beam is right-handed with positive $z$ pointing in the direction of the beam (from right to left in Figure~\ref{fig:TB_foto}), $x$ in the horizontal and $y$ in the vertical direction. The sample is mounted such that columns of the FE chip and rows of the HV2FEI4v2 are oriented vertically.

\subsection{Track Reconstruction}

Two software frameworks are used to reconstruct and analyse the test beam data samples. Track finding and reconstruction is carried out with 
\texttt{EUTelescope}~\cite{Matze}. The telescope tracks and hits from the assembly under test are stored in ROOT ntuples, which are read by the second software framework, \texttt{TBmonII}~\cite{Rieger_Thesis} to run the analysis.\\
The track reconstruction is done in five steps:
\begin{enumerate}
	\item Data conversion:\\
		Raw data from the telescope and DUT DAQ systems are converted into LCIO format. In the process, noisy pixels are identified based on their firing frequency and stored in a database for the following steps.
	\item Clustering:\\
		Individual hits are grouped into clusters based on their spatial distance, with an additional cut on the \lvl of the hits. After 
                having discarded clusters containing pixels that were flagged as noisy, the remaining clusters are stored in a database for further analysis. 
	\item Hitmaking:\\
		Cluster coordinates are derived from all the pixels contained within the cluster. A centre-of-gravity algorithm is used to determine the cluster central position. Correlations between hit positions are used to calculate a rough pre-alignment for the planes.
	\item Alignment:\\
		Pre-aligned hits are used for fitting of preliminary tracks, which then serve as input for alignment in \texttt{MillipedeII}~\cite{MillepedeII}. The resulting alignment constants are stored in a database.
	\item Track fitting:\\
		Tracks are fitted to the aligned hits using a Deterministic Annealing Fitter~\cite{DAF_Havard}, which takes into account multiple scattering. Tracks are required to have hits attached to them on at least five out of the six telescope planes and have $\chi^2\,\leq\,75$.
\end{enumerate}

\subsection{Analysis}

Additional cuts are applied to the ensemble of reconstructed tracks to select only those tracks that are meaningful for the analysis of the DUT.
\begin{itemize}
	\item Tracks are required to have a hit attached to them in the reference plane in order to select tracks that pass through the DUT during its integration time of 400~ns. In the analysis framework, a hit is matched to a track if the distance between the hit coordinate and the extrapolated track position is less than pixel pitch plus 10~$\upmu$m in either direction.
	\item Highest track reconstruction quality is ensured by cutting on $\chi^2\leq 25$ for all reconstructed tracks. 
	\item Tracks that pass through pixels at the edge of the DUT sensor are only regarded in special analyses, but are ignored for general analysis. This is done to suppress possible effects of the electric field at the sensor edge.
	\item Tracks passing through noisy or defect pixels are rejected, since the measured total charge of a cluster containing such pixels might be wrong.
\end{itemize}
Tracks passing these additional cuts are referred to as good tracks and are being used for further analysis.

%

%% file: sections/subpix.tex
\section{Subpixel Decoding} \label{sec:subpixmap} 

To achieve the maximum resolution possible with the HV2FEI4v2, hits in the three subpixels have to be distinguished from each other. To study ways to identify the hit subpixel, test beam measurements were performed with only one subpixel type enabled at a time. The optimised settings for \VNOut{1/2/3} and the \FDAC found in Section~\ref{sec:subpixtune} were used during these measurements.

\begin{figure}[tbph]
\centering
\includegraphics[width=0.75\textwidth]{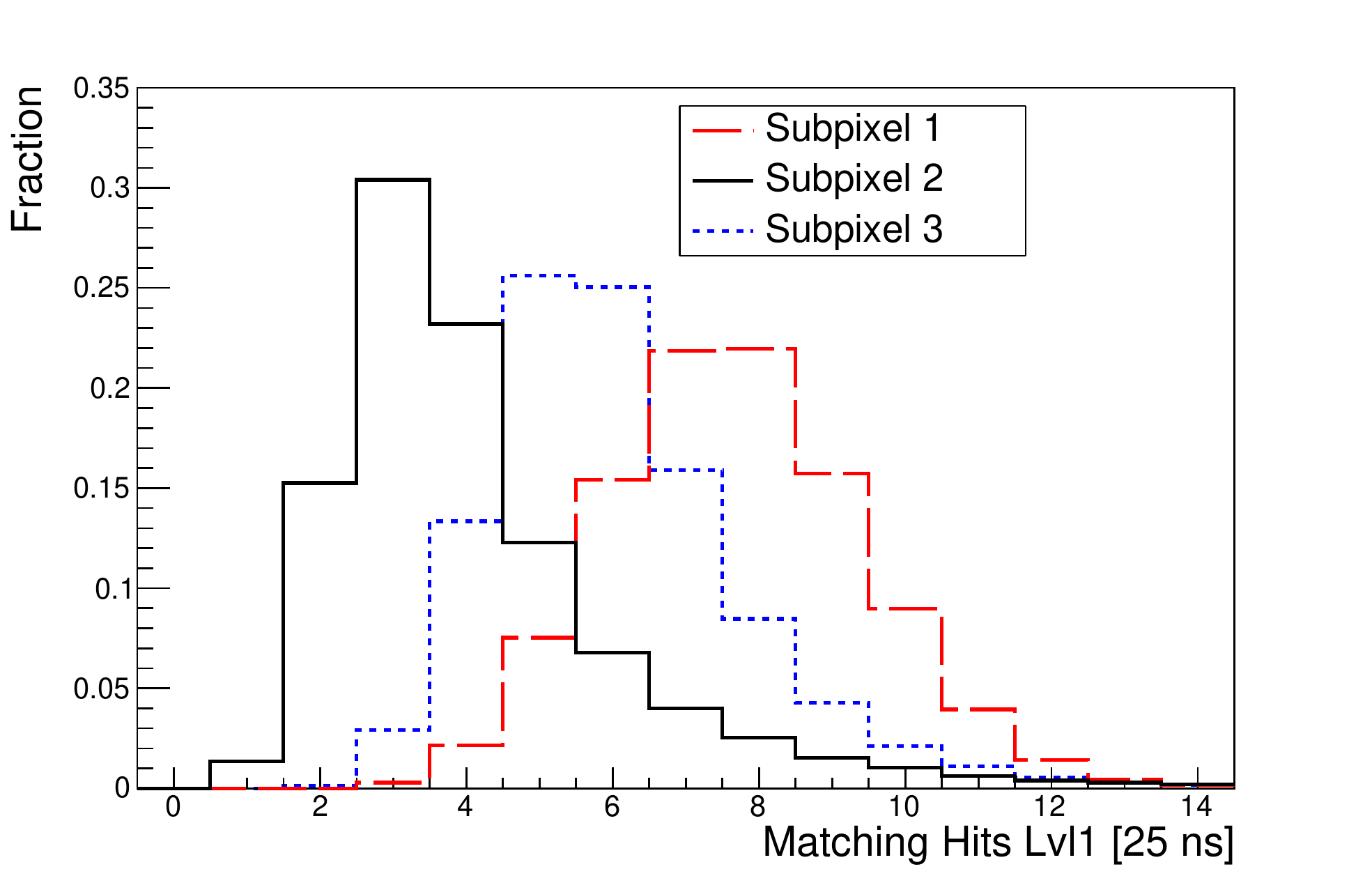}
\caption{\lvl distributions from three separate test beam measurements with only one enabled subpixel type at a time.}
\label{fig:Lvl1}
\end{figure}

The most straight-forward separation of the three subpixel types uses only the \ToT values measured by FE-I4. Since the overlap between the \ToT distributions of the 
subpixel types (see Figure~\ref{fig:HVToT}) prevents an unambiguous distinction, additional variables are studied to obtain larger separation power. 
Figure~\ref{fig:Lvl1} shows the \lvl distributions of the three subpixel types.
The peaks of the \lvl distributions of the three subpixel types differ by two bins from each other. 
This behaviour can be attributed to a combination of the time constant of the coupling capacitor in the HV2FEI4v2 and time walk in the FE-I4 amplifiers.

\begin{figure}[tbph]
\centering
\subfigure[Subpixel 1.\label{fig:LH1}]{\includegraphics[trim=0cm 0cm 0cm 0cm, clip=true,width=0.49\textwidth]{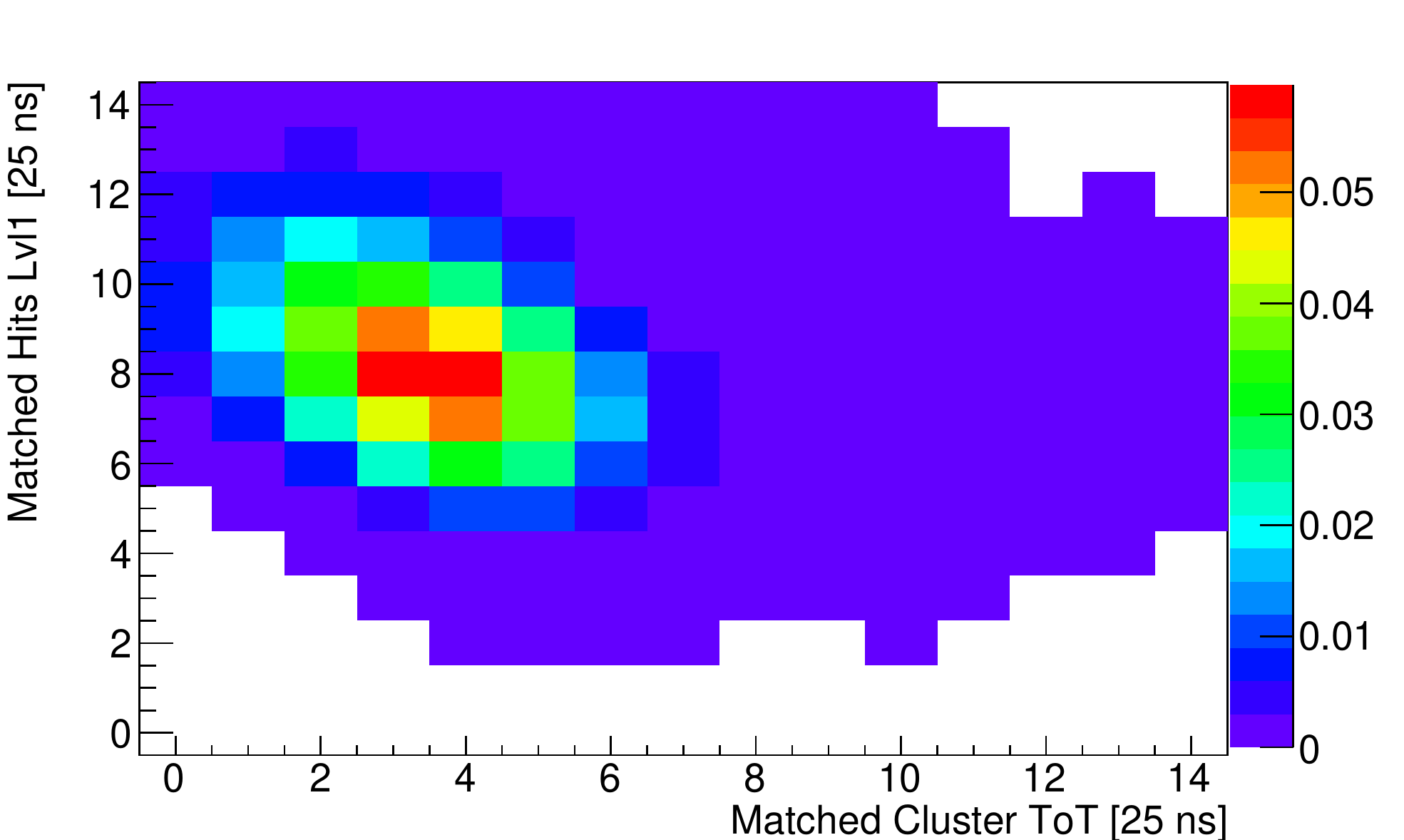}}
\subfigure[Subpixel 2.\label{fig:LH2}]{\includegraphics[trim=0cm 0cm 0cm 0cm, clip=true,width=0.49\textwidth]{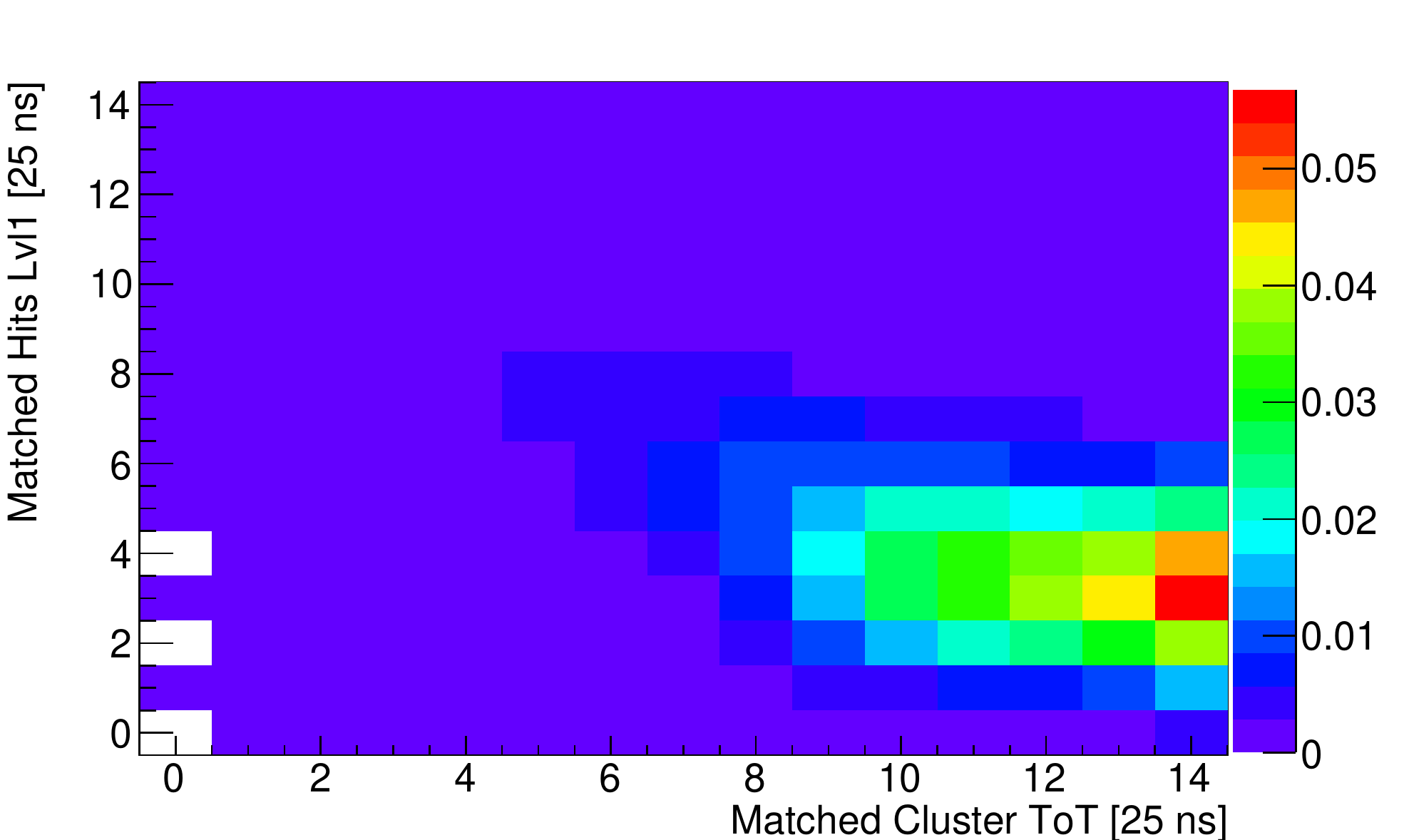}}
\subfigure[Subpixel 3. \label{fig:LH3}]{\includegraphics[trim=0cm 0cm 0cm 0cm, clip=true, width=0.50\textwidth]{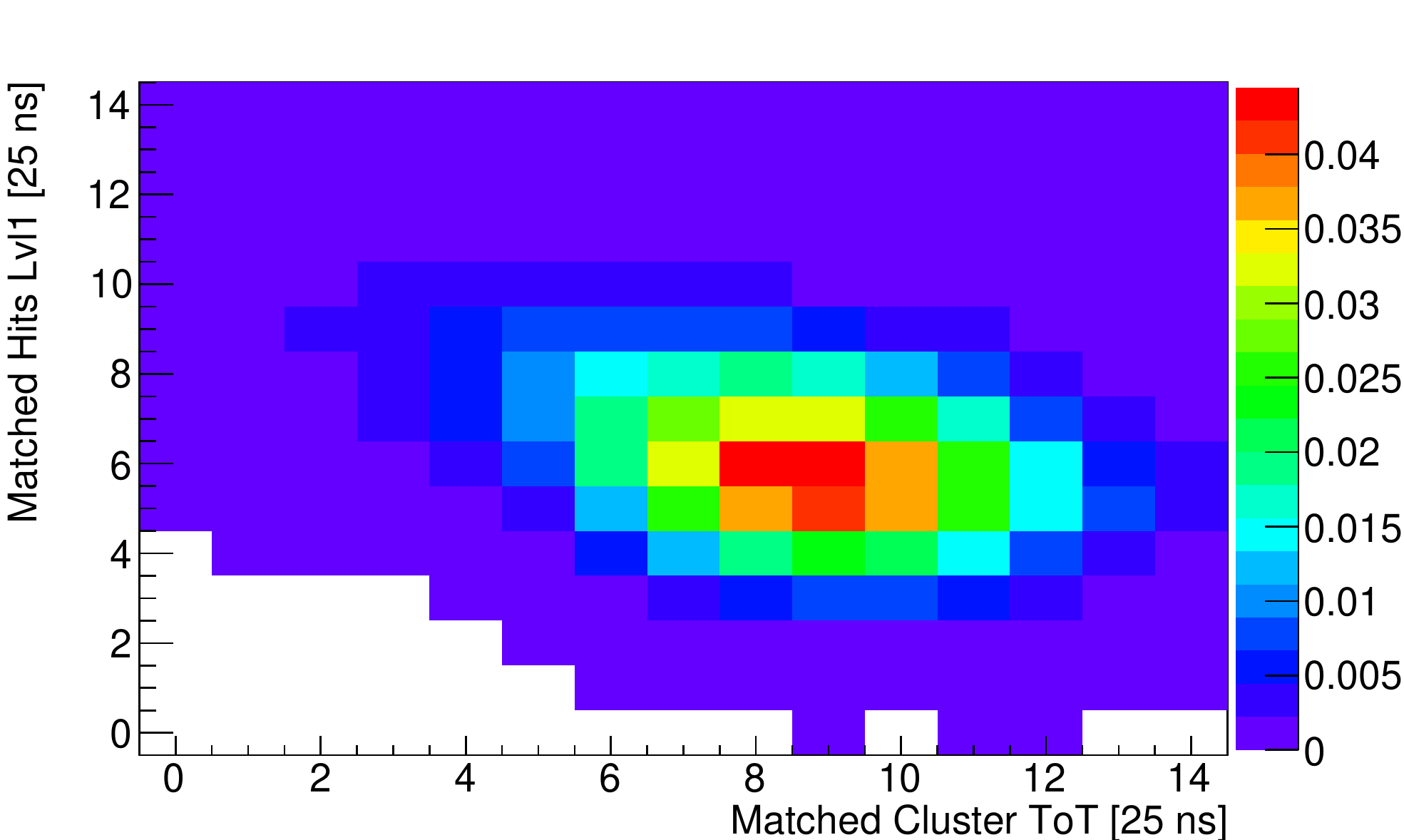}}
\caption{\ToT and \lvl distributions from dedicated test beam measurements with only one subpixel type active at a time. These distributions serve as reference templates for the likelihood.}
\label{fig:LH}
\end{figure}

For the analysis of the test beam measurements the additional information from \lvl is used together with the \ToT information to define a two-dimensional likelihood function for 
each subpixel. Figure~\ref{fig:LH} shows the normalised likelihood distributions for the subpixel types.
These functions were measured in dedicated runs with only one subpixel type active at a time. Since the tuning of both HV2FEI4v2 and FE-I4 read-out chip was not changed, they could be used in subsequent measurements with all subpixels enabled.

%% file: sections/resolution.tex
\section{Resolution Studies} \label{sec:resolution}

Data from test beam measurements with one subpixel type enabled are used to study spatial resolution. 
Due to the layout and arrangement of the HV2FEI4v2 unit cells (see Figure~\ref{fig:subpixel}) the distance between neighbouring active subpixels is large in three of the four directions. Therefore, only clusters with one hit pixel are analysed.

The expected standard deviation is $37.5~\upmu$m in $x$- and $13.8~\upmu$m in $y$-direction, 
given by the quadratic sum of the geometrical resolution 
and the measured telescope pointing resolution of approximately $10~\upmu$m~\cite{Rieger_Thesis}.
The standard  deviations of the measured residual distributions are $\left(35.6\pm 0.1\right)~\upmu$m in $x$-direction and
$\left(15.65\pm 0.06\right)~\upmu$m in $y$-direction. The smaller value for the $x$-direction can be attributed to errors in the cluster position, because
hits close to the edge of a subpixel are treated incorrectly since the neighbouring subpixels are turned off. The larger value for the $y$-direction is caused by a misalignment
during track reconstruction. 

Test beam measurements were carried out with all subpixel types enabled and at a sensor bias voltage of -40~V.
Residual distributions for cluster size one are studied using the likelihood method for subpixel decoding, as described in Section~\ref{sec:subpixmap}.
The standard deviations of the residuals are measured to be $40.9~\upmu$m in $x$ and $15.3~\upmu$m in $y$.
The value in $x$-direction is larger than that obtained with only one subpixel enabled, an effect expected from a wrong assignment
of the subpixel type. Subpixels being mapped to the wrong position cause tails on both sides of the $x$-residual distributions.
The number of mismatched hits was estimated from the tails of the $x$-residual distributions to be approximately~5\%.


%% file: sections/hiteff.tex
\section{Hit Efficiency Studies} \label{sec:hiteff}
The detection efficiency of the HV2FEI4v2 is studied in test beam measurements with all subpixel types enabled. It is defined as the fraction of good tracks matched to hits on the HV2FEI4v2 within a matching radius of pixel pitch plus 10~$\upmu$m around the extrapolated track position, divided by the total number of good tracks passing through the DUT.
Data was taken at room temperature with sensor bias voltages of -30~V to -80~V, with most results based on measurements at -60~V.
%
The errors induced by a subpixel mismatch rate of approx.~5\% (see Section~\ref{sec:resolution}) propagates to the hit efficiency. 
Hence, the relative uncertainty on the hit efficiency is dominated by the mismatch rate.

The mean hit efficiencies and the standard deviation $\sigma$ measured at the minimum stable external threshold voltage of 0.89~V
are presented in Table~\ref{tab:3SP_Eff}.
\begin{table}[htb]
\centering
\begin{tabular}{c|c|c|c|c|c|c|c}
\multicolumn{2}{c|}{Module average} &  \multicolumn{2}{c|}{Subpixel 1} & \multicolumn{2}{c|}{Subpixel 2} &
 \multicolumn{2}{c}{Subpixel 3} \\
Efficiency & $\sigma$ & Efficiency & $\sigma$ & Efficiency & $\sigma$ & Efficiency & $\sigma$\\
\hline
 72\% & 21\% & 55\% & 28\% & 81\% & 9\% & 80\% & 9\% \\
\end{tabular}
\caption{Mean hit efficiencies and their standard deviation for a test beam measurement with all subpixel types enabled at 
a sensor bias voltage of -60~V.}
\label{tab:3SP_Eff}
\end{table}
The mean hit efficiency of subpixel type~2 and~3 is approximately~80\%, the maximum pixel efficiency being $\geq 95\%$. The hit efficiency distribution of subpixel 
type~1 is very broad and the mean hit efficiency is lower than for the other subpixel types, which lowers the average hit efficiency across the whole sensor.
%
%
%
This effect is attributed to two separate issues. First, the hit efficiency of subpixel type~1 is measured to be intrinsically lower than for the others. Second, the 
low \ToT response of subpixel type~1 together with the high HV2FEI4v2 comparator threshold\footnote{The actual comparator threshold is determined by the difference of the external 
threshold voltage to the baseline voltage~\cite{IvanCCPD2}, the latter being set to 0.8~V for all measurements. This has to be compared with a signal voltage of approximately 100~mV measured for a Fe55 
line~\cite{IvanHVCMOS} which generates $1660e$ of charge in silicon, while the most probable charge deposited by a MIP-like particle
in $10~\upmu$m of silicon is approximately $1000e$.} further reduces the efficiency.

\begin{figure}[htb]
\vspace{1.5cm}
\centering{
\subfigure[\VNOut{3}.\label{fig:3SP_EffGeo15}]{\includegraphics[trim=0cm 0cm 0cm 0cm, clip=true, width=0.495\textwidth]{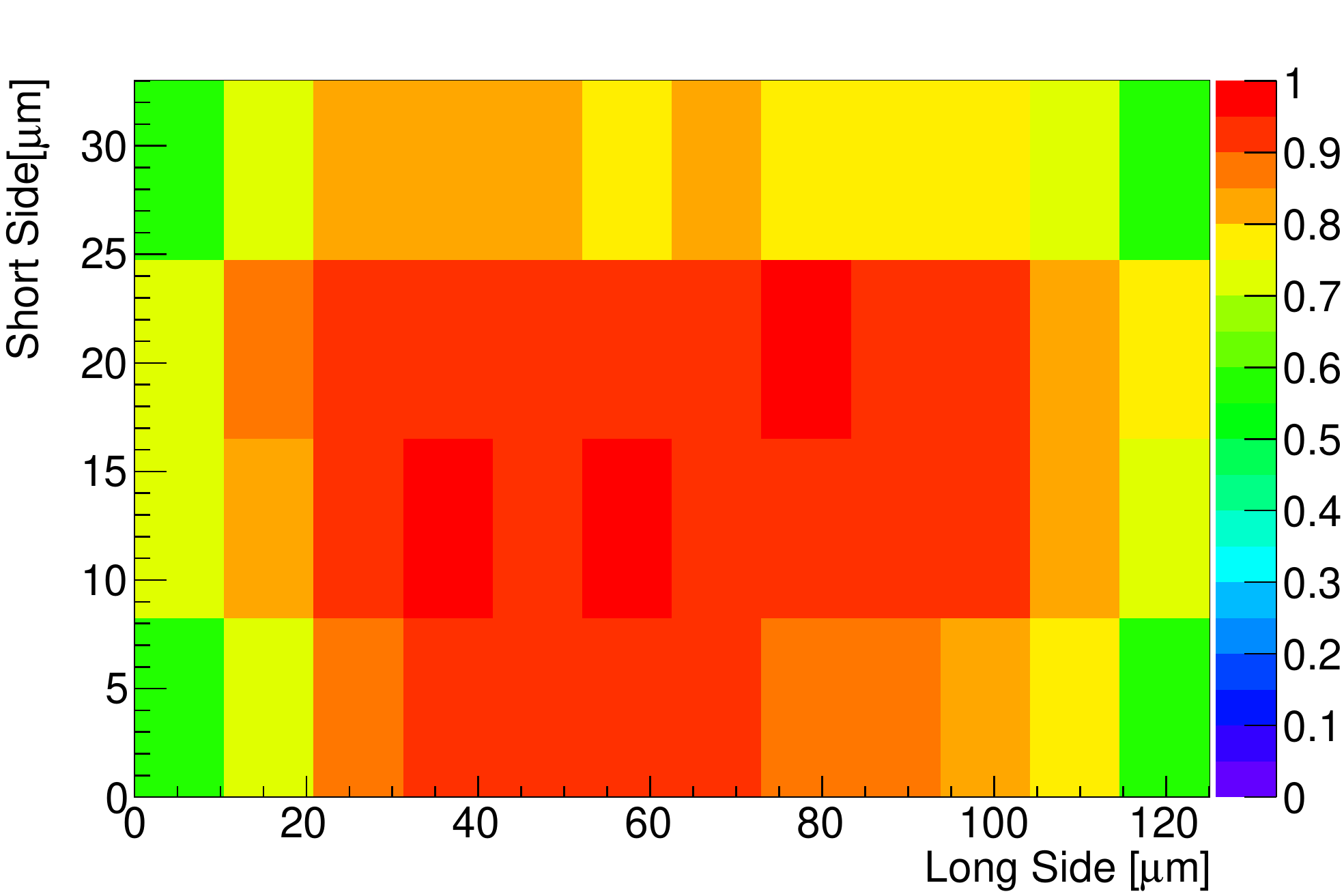}}
\subfigure[\VNOut{3}*.\label{fig:3SP_EffGeo18}]{\includegraphics[trim=0cm 0cm 0cm 0cm, clip=true,width=0.495\textwidth]{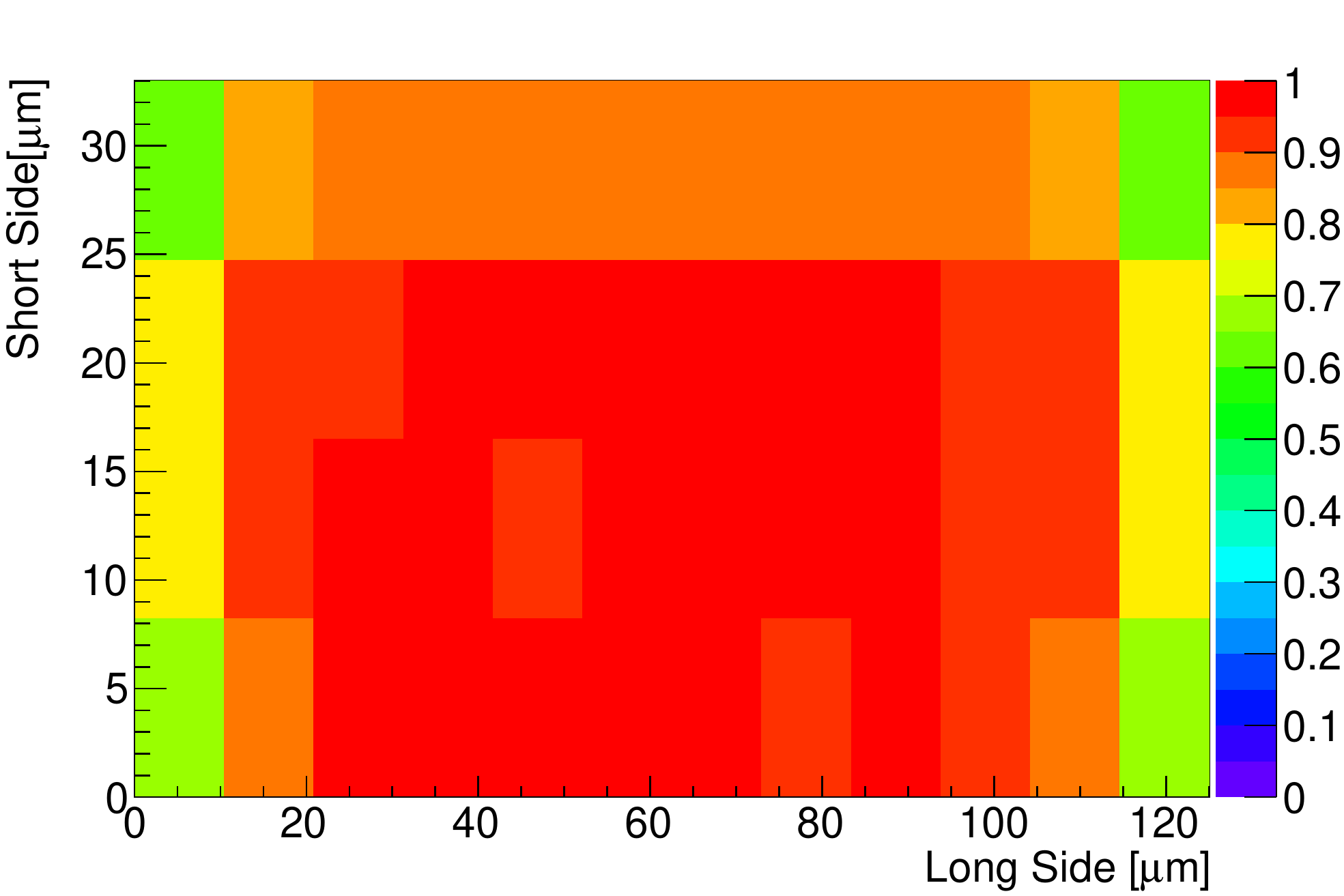}}
\subfigure[\VNOut{2}*.\label{fig:3SP_EffGeo17}]{\includegraphics[trim=0cm 0cm 0cm 0cm, clip=true,width=0.495\textwidth]{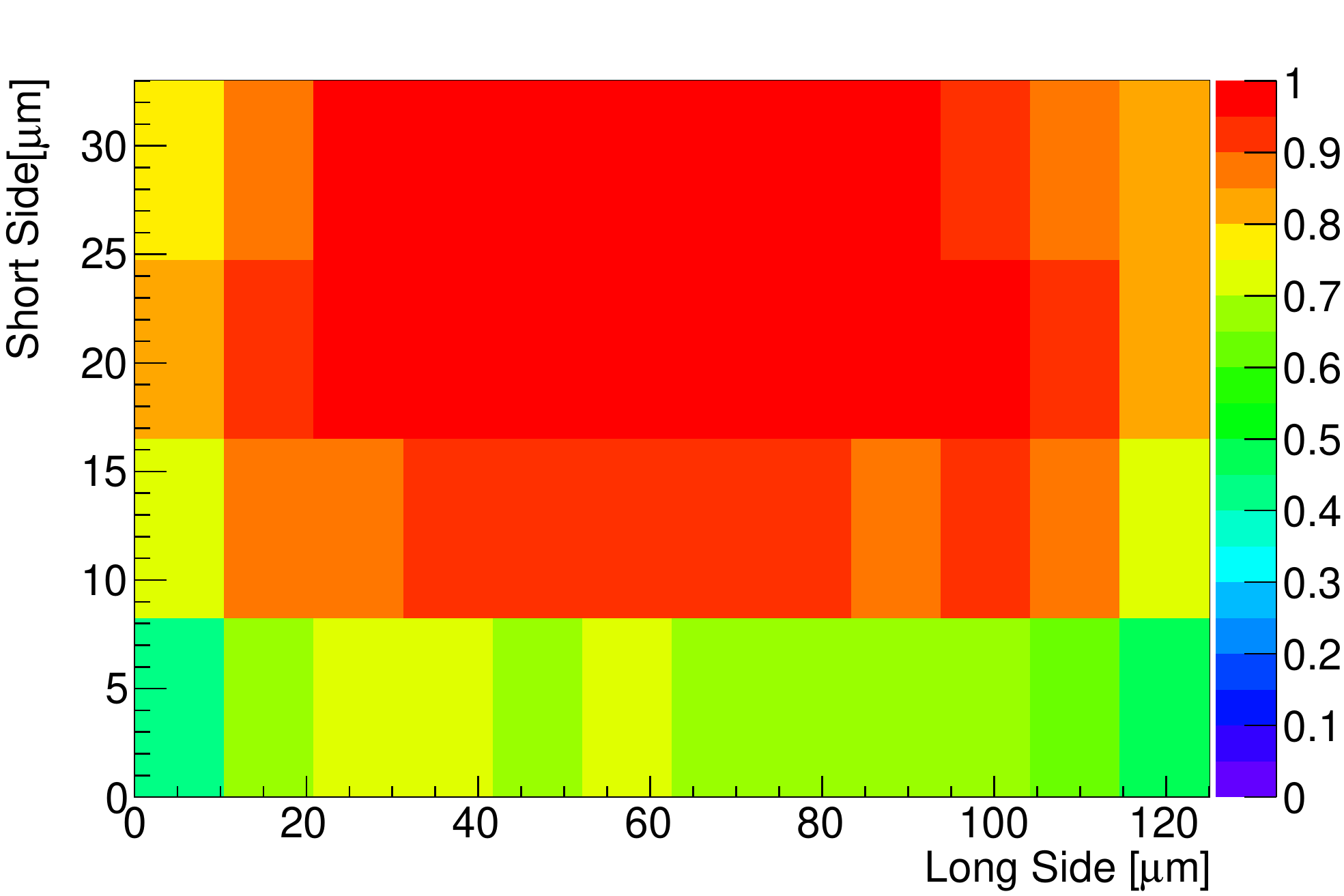}}
\subfigure[\VNOut{2}.\label{fig:3SP_EffGeo14}]{\includegraphics[trim=0cm 0cm 0cm 0cm, clip=true,width=0.495\textwidth]{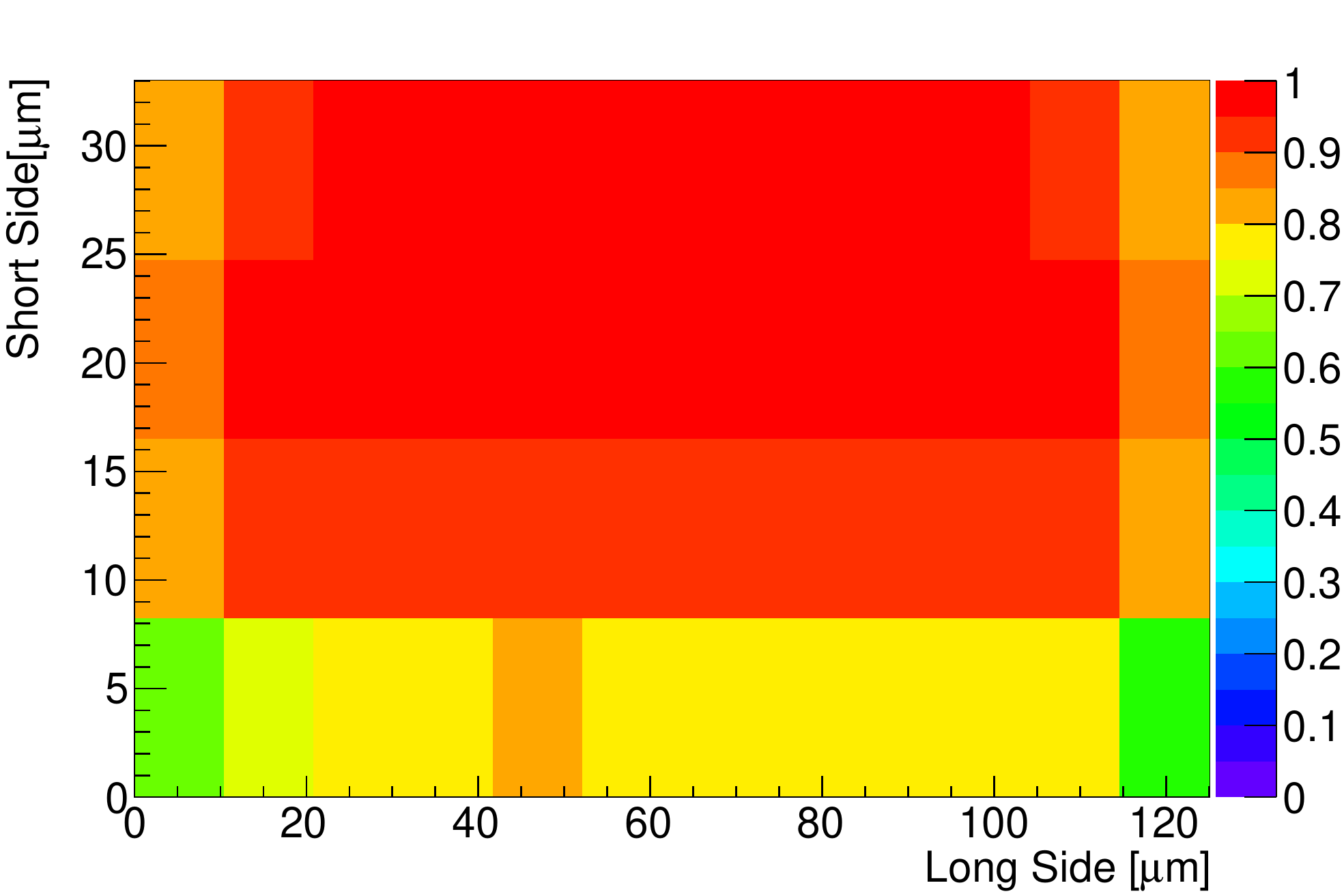}}
\subfigure[\VNOut{1}.\label{fig:3SP_EffGeo13}]{\includegraphics[trim=0cm 0cm 0cm 0cm, clip=true,width=0.495\textwidth]{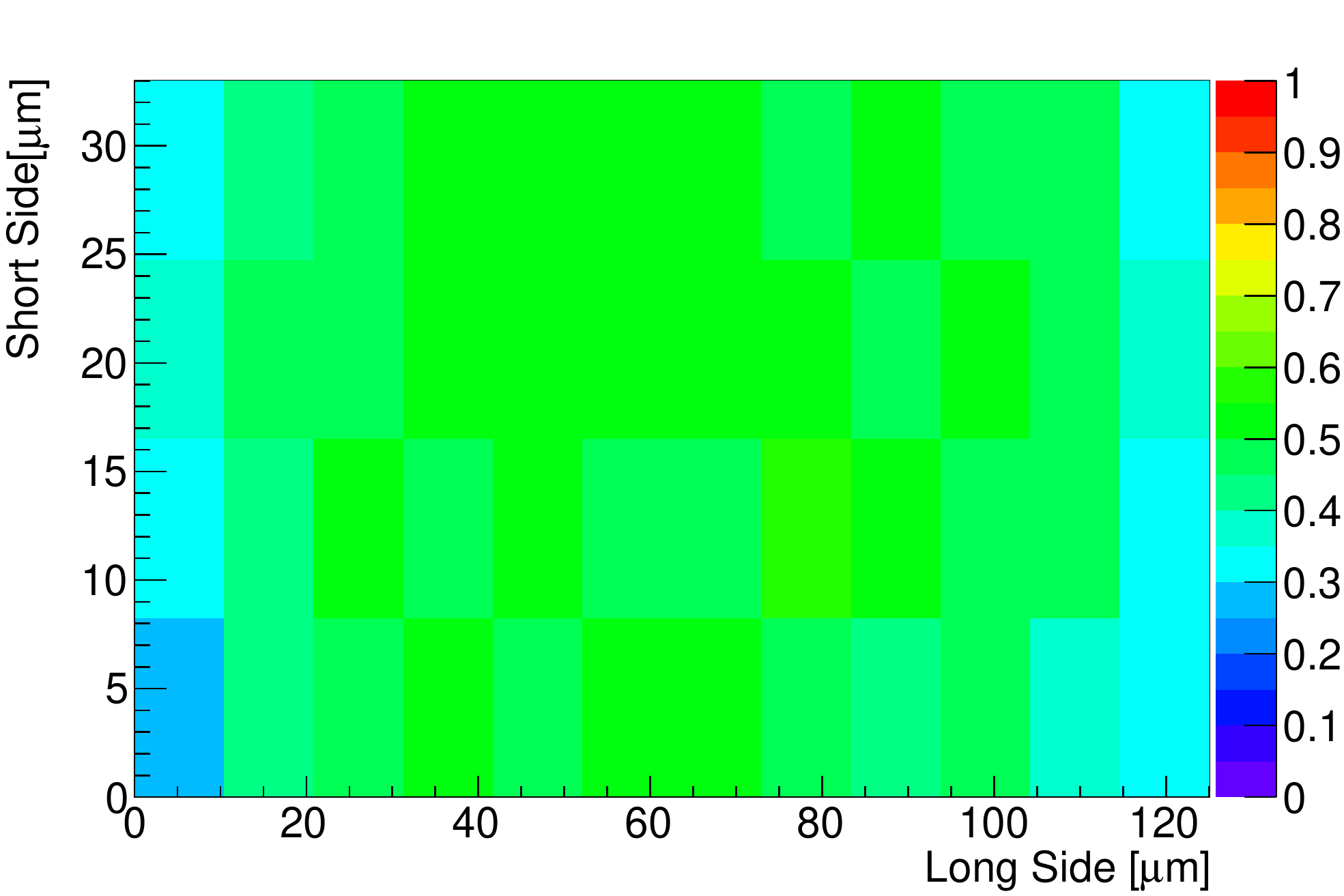}}
\subfigure[\VNOut{1}*.\label{fig:3SP_EffGeo16}]{\includegraphics[trim=0cm 0cm 0cm 0cm, clip=true,width=0.495\textwidth]{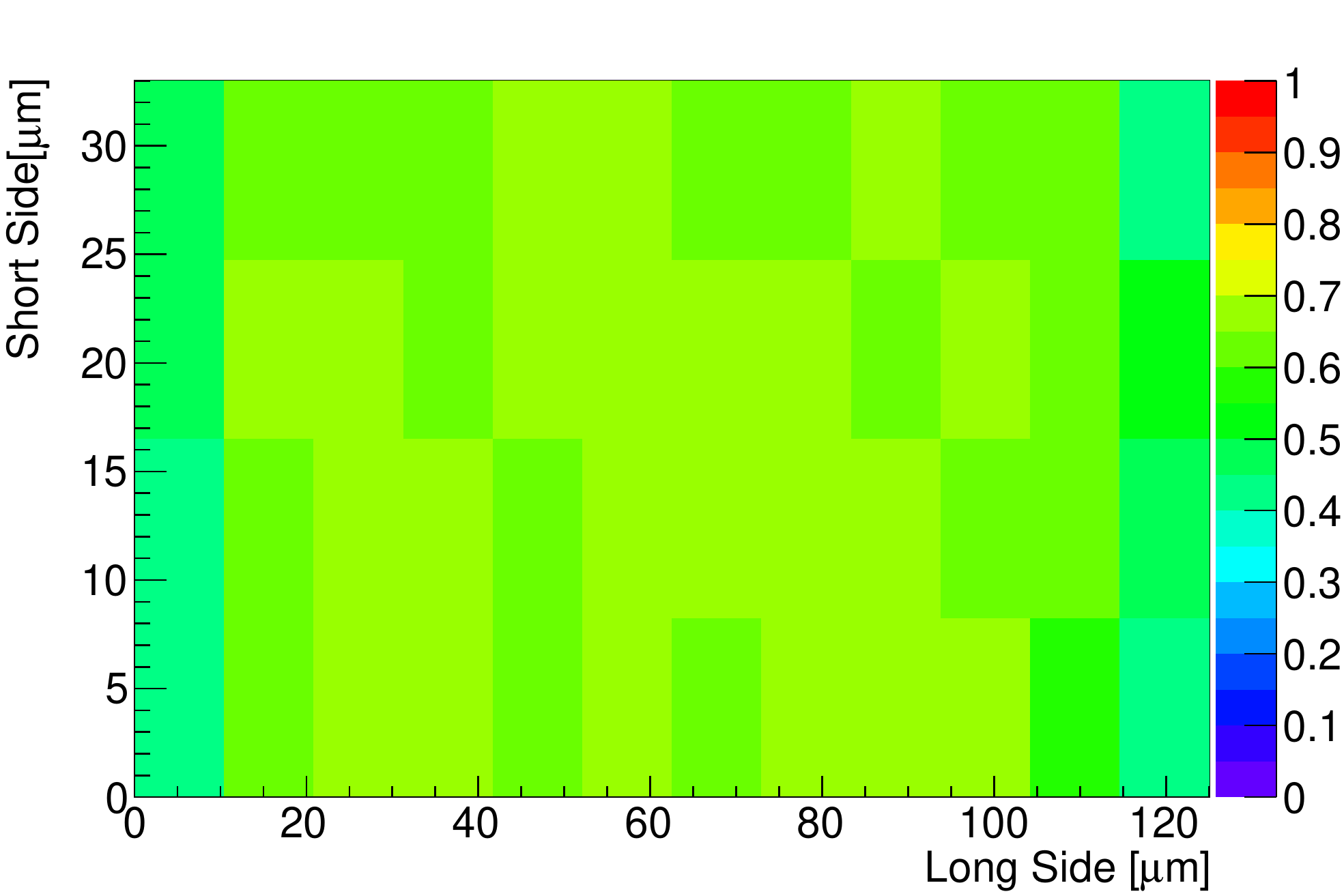}}
\caption{In-pixel hit efficiency maps for the 6 different subpixel geometries. All subpixel types were enabled. The external threshold voltage is 0.89~V and 
the sensor bias voltage -60~V. The colour scale ranges from 0 to 1.} 
\label{fig:3SP_EffGeo}}
\vspace{1cm}
\end{figure}

The average in-pixel efficiency plotted separately for the six pixels of one unit cell is shown in Figure~\ref{fig:3SP_EffGeo}. To increase statistics, pixel hits and tracks from all unit cells are projected into one cell, taking the cell orientation into account.
The hit efficiency in the centre of the subpixel reaches values in excess of 90\% for subpixel types 2 and 3, decreasing towards the edges of the subpixel due to charge sharing.
This, together with the low efficiency of subpixel type 1, creates the asymmetry of the efficiency at the long edges of subpixel types 2 and 3.

\begin{figure}[htbp]
\centering
\subfigure[External threshold voltage. \label{fig:3SP_Eff_vs_Thr}]{\includegraphics[width=0.45\textwidth]{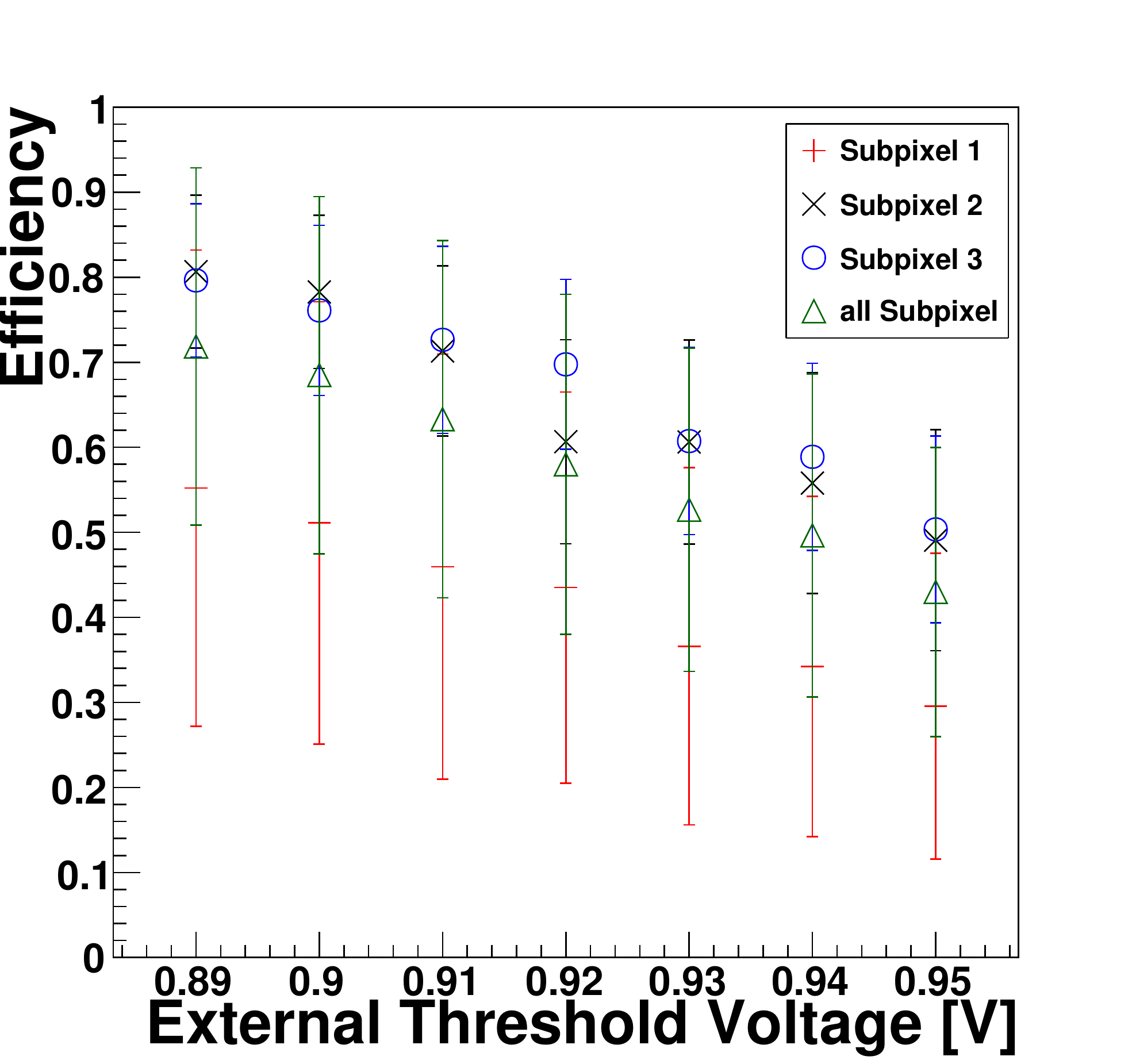}}
\hfill
\subfigure[Sensor bias voltage. \label{fig:3SP_Eff_vs_HV}]{\includegraphics[width=0.45\textwidth]{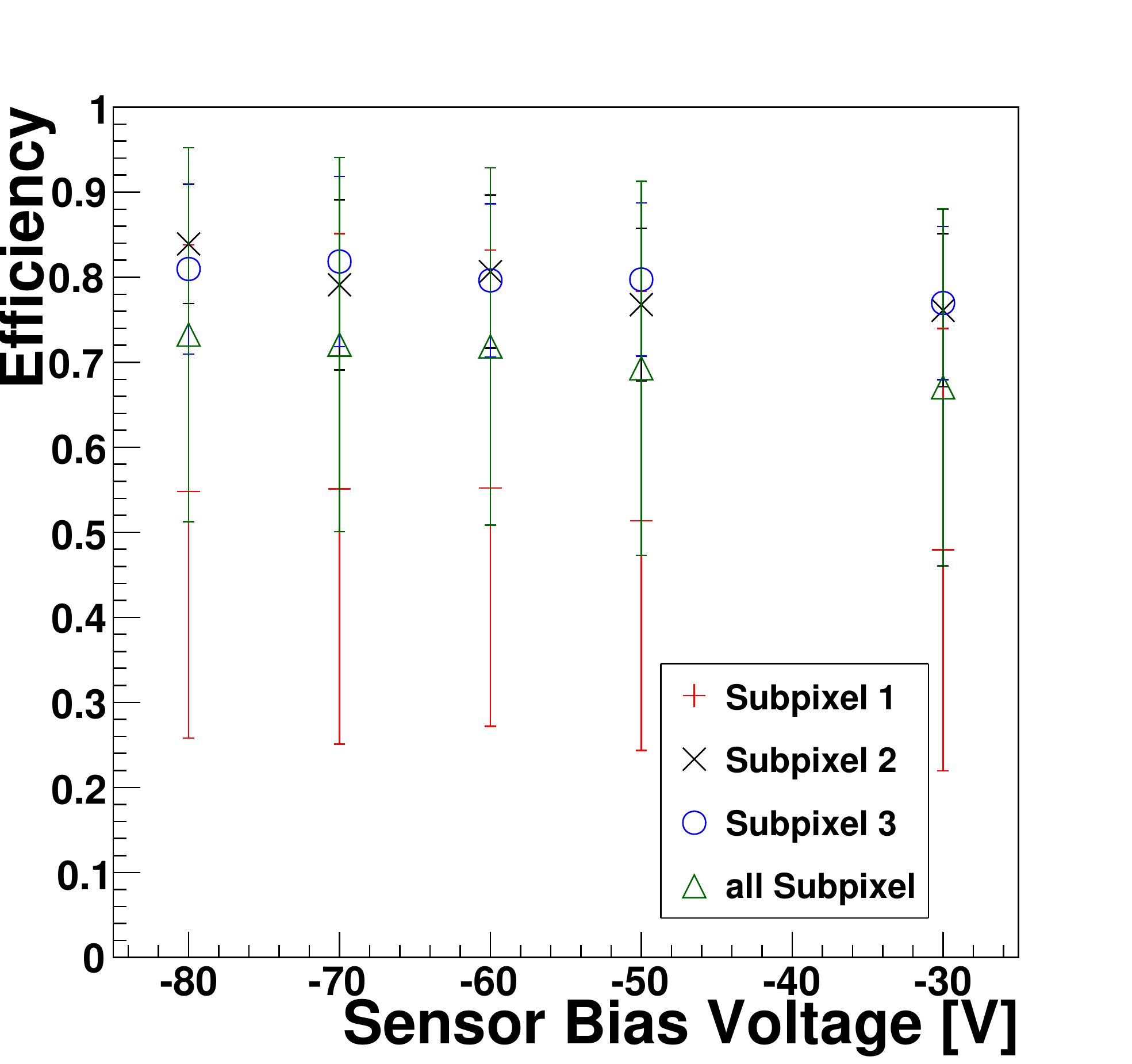}}
\caption{Hit efficiency as a function of the external threshold voltage at a sensor bias voltage of -60 V and the sensor bias voltage for an external threshold 
voltage of 0.89 V. All subpixel types are enabled. The error bar shows the standard deviation of the efficiency distribution.}
\end{figure}

The hit efficiency as a function of the external threshold voltage is shown in Figure~\ref{fig:3SP_Eff_vs_Thr} for external threshold voltages between 0.89~V 
and 0.95~V. It decreases with increasing external threshold voltage, as expected from the corresponding increase in HV2FEI4v2 comparator threshold discussed 
in Section~\ref{sec:compthresh}.

\begin{figure}[ht]
\centering
\subfigure[Sensor bias voltage -60~V. \label{fig:3SP_1DEff_60}]{\includegraphics[trim=0cm 0cm 0cm 0cm, clip=true,width=0.5\textwidth]{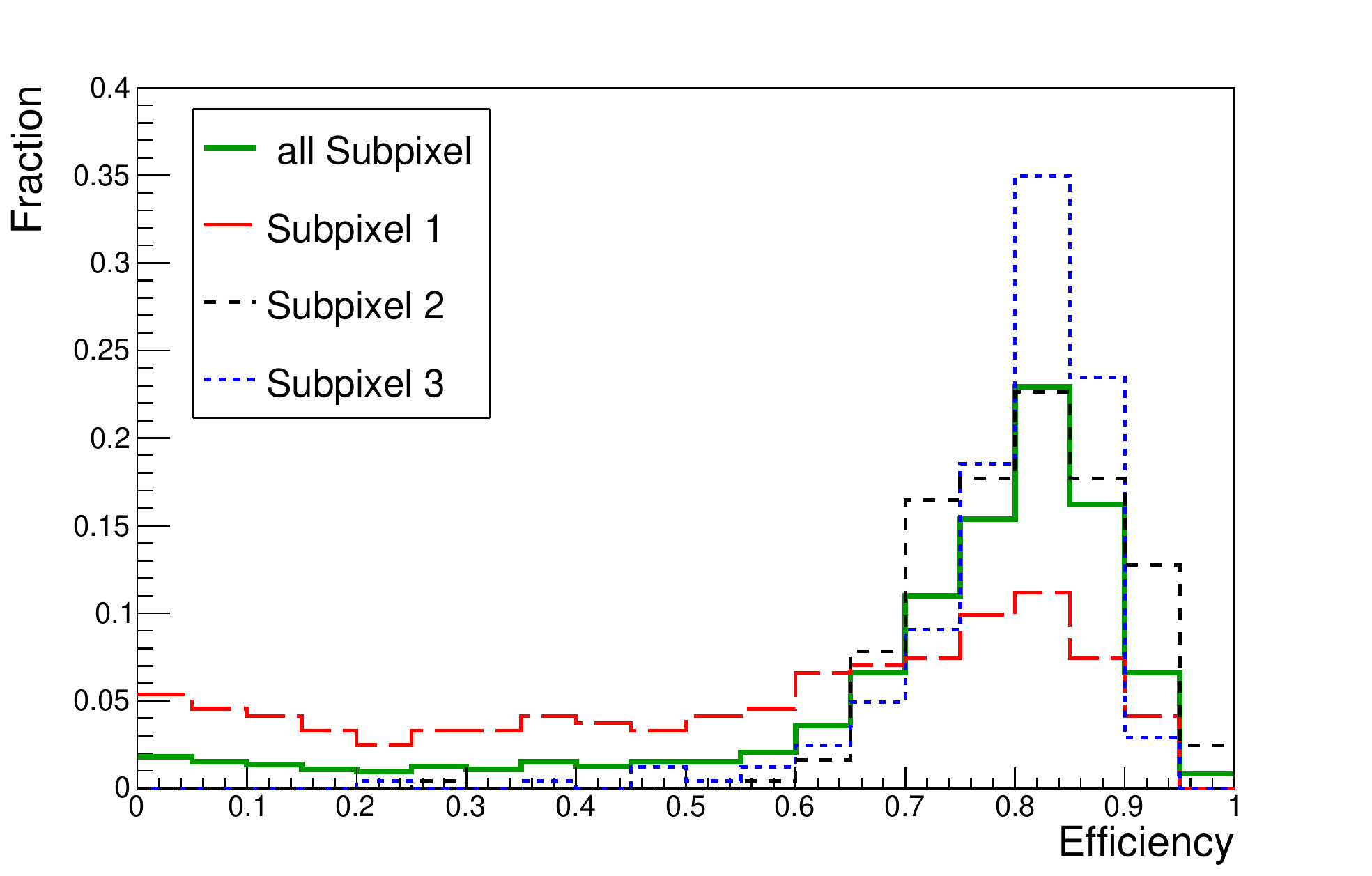}}
\subfigure[Sensor bias voltage -80~V. \label{fig:3SP_1DEff_80}]{\includegraphics[trim=0cm 0cm 0cm 0cm, clip=true,width=0.48\textwidth]{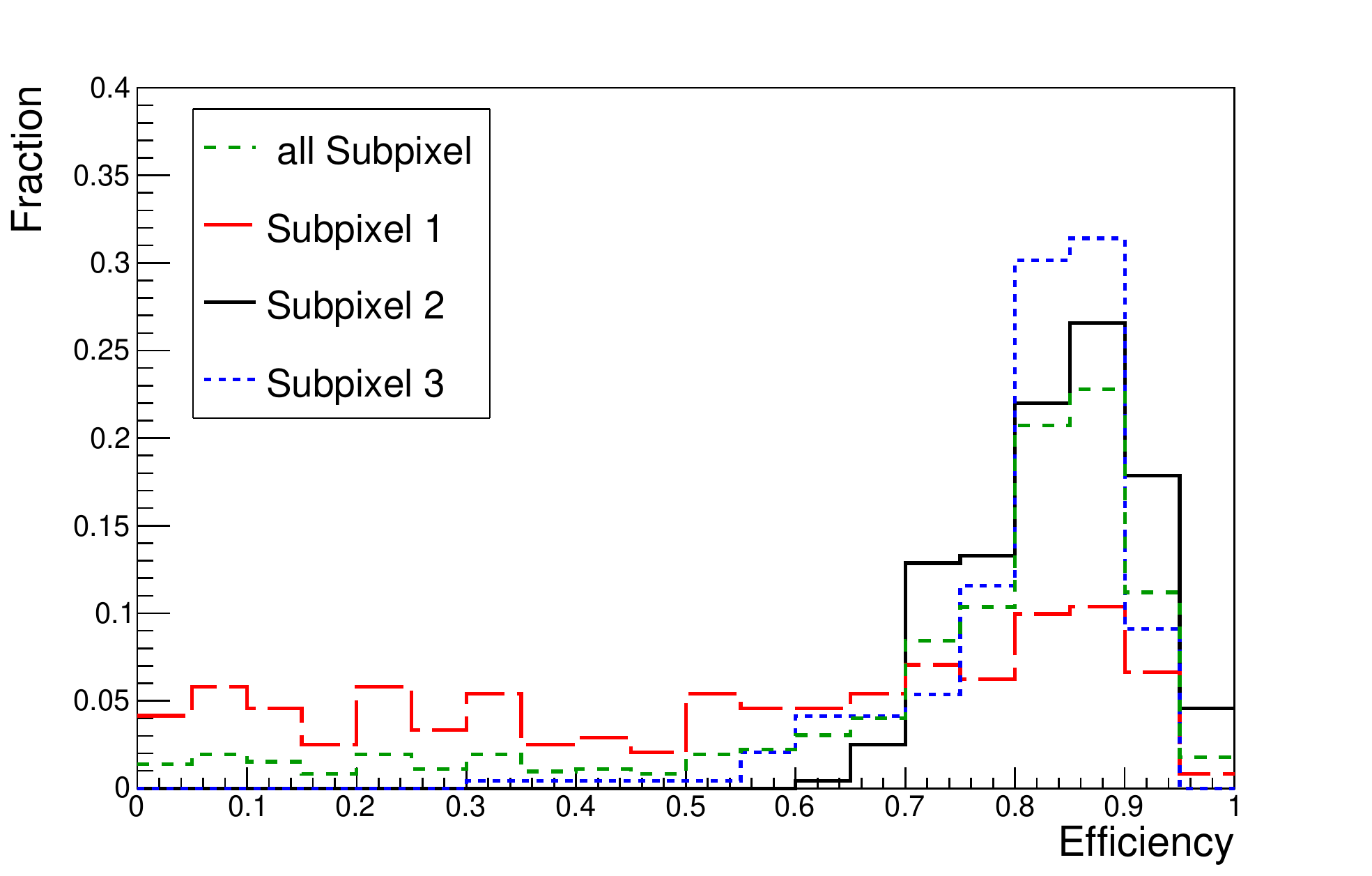}}
\caption{Hit efficiency distribution for a test beam measurement with all subpixel types enabled for different sensor bias voltages. The external threshold voltage is 
0.89~V.}
\end{figure}

Figure~\ref{fig:3SP_Eff_vs_HV} shows the hit efficiency as a function of the sensor bias voltage for bias voltages between -30~V and -80~V. 
Again, the large standard deviation of subpixel type 1 can be seen. No significant increase in the mean hit efficiency for a decreasing sensor bias voltage 
is observed. However, the hit efficiency distributions of -60~V (see Figure~\ref{fig:3SP_1DEff_60}) and -80~V (see Figure~\ref{fig:3SP_1DEff_80}) exhibit a 
shift of the peak position of the distribution. This indicates an increase of the efficiency with the sensor bias voltage as expected from faster charge 
collection due to an increased drift field.

%% file: sections/conclusion.tex
\section{Conclusion}

This paper describes the characterisation of a CCPD module with HV2FEI4v2 sensor, which is implemented in the AMS 180~nm high-voltage process.
The investigated CCPD module connects every FE-I4 pixel to three HV2FEI4v2 subpixels, 
the position of which is encoded in the output signal amplitude. 

Several HV2FEI4v2 parameters were studied, including adjustment of 
the \ToT response of the FE-I4 to the sensor signal for a successful 
subpixel mapping. The parameter range in which a significant change of the output voltage is possible was found to be small, which makes 
subpixel decoding a challenging task. The challenge is met by a subpixel mapping with a likelihood based on \ToT and \lvl.


Data from test beam measurements were analysed with all subpixel types enabled  and subpixel mapping based on the likelihood method.
Residual distributions show a mismatch rate of 5\%.
The mean hit efficiency was measured to $72\%$ with a large standard deviation of 
$21\%$. The mean hit efficiency of subpixel types 2 and 3 is approximately $80\%$. The measurement of the in-pixel efficiency showed that 
the hit efficiency in the centre of the pixel is higher than at the edges due to charge sharing.

The HV2FEI4v2 is a very early prototype towards a CCPD module for the ITk. The feasibility of the subpixel de- and encoding concept
was successfully demonstrated, resulting in an improved pixel resolution compared to that of the larger pixels of the read-out chip. 
The measured mean hit efficiency of the module of $72\%$ is clearly below ITk specifications. This feature can be addressed
by reducing the HV2FEI4v2 comparator threshold.